# Scalable Online Survey Framework: from Sampling to Analysis

Weitao Duan
LinkedIn Corporation
950 W. Maude Ave
Sunnyvale, CA, 94085, USA
wduan@linkedin.com

Qian Weng
LinkedIn Corporation
950 W. Maude Ave
Sunnyvale, CA, 94085, USA
qiweng@linkedin.com

Rogier Verhulst
LinkedIn Corporation
950 W. Maude Ave
Sunnyvale, CA, 94085, USA
rverhulst@linkedin.com

Ya Xu
LinkedIn Corporation
950 W. Maude Ave
Sunnyvale, CA, 94085, USA
yaxu@linkedin.com

## ABSTRACT

With the advancement in technology, raw event data generated by the digital world have grown tremendously. However, such data tend to be insufficient and noisy when it comes to measuring user intention or satisfaction. One effective way to measure user experience directly is through surveys. In particular, with the popularity of online surveys, extensive work has been put in to study this field. Surveys at LinkedIn play a major role in influencing product and marketing decisions and supporting our sales efforts. We run an increasing number of surveys that help us understand shifts in awareness and perceptions with regards to our own products and also to peer companies. As the need to survey grows, both sampling and analysis of surveys have become more challenging. Instead of simply multiplying the number of surveys each user takes, we need a scalable approach to collect enough and representative samples for each survey analysis while maintaining good user experience. In this paper, we start with discussions on how we handle multiple email surveys under such constraints. We then shift our discussions to challenges of in-product surveys and how we address them at LinkedIn through a survey study conducted across two mobile apps. Finally, we share how in-product surveys can be utilized as monitoring tools and connect surveys with A/B testing.

## 1 INTRODUCTION

With the advancement in technology, raw event data generated by the digital world have grown tremendously. Behavioral analytics has been widely used to leverage such data to understand how consumers act, and to predict their actions in the future. However, observational user behavior data is well known to be noisy as they are subject to many external, unknown factors, such as each user's personal preference and seasonal effects. Such data also tend to be insufficient when it comes to measuring user intention or satisfaction. They help answer the questions of "how," but not the questions of "why."

One way to measure user experience directly is through surveys. Survey is a common way to gather information about the population of interest from a sample of individual units. It has been widely used in all kinds of research fields, e.g. marketing [1], psychology [2], health care [3] and sociology [4]. Big Internet companies such as Facebook [5,6], Google [7] and LinkedIn all heavily leverage online user survey to complement their rich user behavior data. Survey data offers users' feedbacks directly and can be used to address specific questions on populations we are interested in.

Online survey has become more and more popular, because it is one of the most cost-effective modes of survey research compared with paper surveys and phone surveys [15]. A side effect for online surveys is that the response rates are significantly lower than phone or mail surveys. While mail surveys have a 4.4% average response rate, online surveys, on average, have merely a 0.75% response rate [14]. There are mainly two channels of conducting surveys online: in-product surveys and email surveys. In-product surveys are shown when users are engaged with the product, best at reflecting user feedback at the moment and at capturing user's attention. On the other hand, email surveys are better for longer questionnaires but the response rates are usually lower than in-product surveys. At LinkedIn, we conduct both email and in-product surveys. Both surveys complement each other and allow us to gather feedback from different angles.

As we increase the number of surveys we conduct, it becomes more apparent that we need a framework that can scale properly. Different from observational studies, a survey interrupts users' regular activity flow. It is crucial to maintain the same number of surveys a user receives while we scale up the survey programs. As a result, surveying all units in the population in many scenarios is infeasible, especially when there are multiple on-going survey programs competing for users. We have to carefully design the sampling process to optimize across all surveys, and to ensure different surveys do not interfere with each other and that samples are representative. If sampled units favor one particular group, results extrapolated from the sample are biased. Sampling bias is one of the major sources for errors in survey studies. Online surveys tend to have additional sampling challenges. First, samples are collected continuously but the population and users available for survey can change over time. This particularly creates challenge when we want to compare survey results across time. Second, it is often needed to ask for user feedback after they perform certain actions on the site. Such trigger-based surveys require sampling to be done in a timely fashion to be effective. Lastly, with the scale of survey programs at LinkedIn, human management becomes inefficient and error-prone. To overcome the above sampling challenges, we have designed and implemented a sampling framework (details in Section 4) that automatically conducts multiple surveys in a consistent and continuous manner.

There is another important type of bias, nonresponse bias, which is especially prevalent among online surveys. Not everyone responds to a survey, and users who respond can be quite different from users who do not respond. Such difference tends to be larger for online surveys because their response rates tend to be lower [14]. The nonresponse bias makes it difficult to compare results from different survey studies, which is often needed as we couple in-product surveys closely with product development. For example, when a product gets a massive overhaul, in addition to A/B testing the new version, we would also want to survey users for direct feedback. However, users who do not like changes are more likely to respond to surveys in the new version than in the old version. In Section 5, we go over these challenges with in-product surveys through a complex case study where we compare survey results from LinkedIn's old vs. new flagship apps.

The contribution of the paper is as follows:
- We propose a novel sampling framework that is able to scale with increasing number of survey programs while protecting user experience.
- We demonstrate with a real example that it is crucial to reduce bias when applying in-product surveys to measure product changes, and the techniques to achieve it.
- The continuous sampling methodology leverages existing A/B systems and can be easily adopted by practitioners across the industry.
- We share learning and insights from solving many practical problems people run into when conducting surveys.

This paper is organized as follows. We start with literature review in Section 2, and then overview how survey serves LinkedIn in general in Section 3. In Section 4, we introduce a sampling framework that addresses the survey challenges in Internet businesses like LinkedIn. In Section 5, we discuss the approaches we adopt to compare survey results from two in-product surveys, then extend the email survey sampling framework to in-product surveys and connect surveys with A/B testing. We conclude with future work directions in Section 6.

## 2   LITERATURE REVIEW

In this section, we review the existing sampling and analysis methods in survey studies.

The major components of survey methodology [8] have been studied in many literatures. In summary, to conduct a survey we need to identify 1) the population of interest and the sampling frame 2) sampling strategy for how to gather an unbiased and representative sample of individual units 3) the associated survey data collection techniques, such as questionnaire construction and methods for improving the number and accuracy of survey responses, 4) the data analysis strategy, which often includes instruments or procedures that correct the non-response bias.

To measure survey quality, some researchers [9, 10, 11] categorize biases into sampling bias and nonsampling bias. Sampling bias is due to selecting a biased sample to study the whole population. [12] summarizes some common sampling techniques to reduce or eliminate sampling bias. Nonsampling bias is a result of mistakes and/or system deficiencies. It includes all errors that can be made during data collection and data processing, such as coverage, nonresponse, measurement, and coding error (see also Chapter 22 in [13]). Online surveys particularly suffer from nonresponse bias because of its low response rate. Weighting adjustment [16] and propensity modeling [17, 18, 19] are the two commonly used methods to adjust nonresponse bias.

Many companies conduct surveys on third-party platforms. Google Survey[20], Facebook Survey[21] and SurveyMonkey [22] are among the most popular platforms for running online survey programs. As far as we know, all these tools focus on managing one survey at a time. For companies like LinkedIn, running a large number of survey programs simultaneously while maintaining good user experience becomes infeasible if we use these existing platforms. To address the scalability issue, we develop an in-house online survey framework.

## 3   NPS SURVEYS AT LINKEDIN

Surveys at LinkedIn play a major role in influencing product and marketing decisions and in supporting our sales efforts. Insights generated from surveys augment our understanding of not only what members do on LinkedIn through site traffic data but also why. We conduct many types of surveys at LinkedIn to help answer how members react to new product features, how loyal they are to the services and what their general perceptions are. For example, we run a number of large scale tracking surveys that help us understand shifts in awareness and perceptions vis-à-vis peer companies as a result of changes in product, marketing and marketplace dynamics. Among all the surveys we do, NPS survey is the most widely used across different products at LinkedIn.

In a NPS survey, we measure "Net promoter score" (NPS) - a metric that was first pioneered by Frederick Reicheld and Satmetrix in 2003 [23] as the one metric that best predicts profitability. NPS is calculated based on responses on the key question: How likely is it that you would recommend our company/product to a friend or colleague? The response is based on a 0-10 scale where 10 means "extremely likely."

Respondents are then grouped as follows:
- **Promoters** (score 9-10) are loyal customers who will keep using the LinkedIn products and refer others.
- **Passives** (score 7-8) are generally satisfied but unenthusiastic customers who are vulnerable to competitive products.
- **Detractors** (score 0-6) are unhappy customers who have high rates of churn and can damage the brand and impede growth through negative word-of-mouth.

NPS is the percentage of promoters minus the percentage of detractors with a range from -100 to 100.

NPS has been widely adopted by more than two thirds of Fortune 1000 companies to measure customer satisfaction. It is a key metric for determining the health of the relationship that members have with our products and services, and to predict future growth in member engagement and revenue. NPS surveys help segment our members in terms of their needs and formulate competitive strategies to stay ahead of competitors and reduce churn. Because of its importance at LinkedIn and in industry at large, we discuss the survey sampling and analysis methodology in the context of NPS surveys. The same methodology applies to other surveys in general.

## 4 EMAIL SURVEYS

As discussed in Section 3, at LinkedIn we care about overall user satisfaction as well as feedback on individual product areas. Therefore, two themes of email surveys are conducted to answer these key questions. In this section, we focus on the the sampling methods. Bias adjustment can be done similarly as in-product surveys, which will be discussed in depth in Section 5. We start by introducing the two email survey themes and sampling constraints. Then we dive into each theme and discuss how we manage multiple survey programs through a scalable sampling framework.

### 4.1 Multiple Survey Programs at LinkedIn

In this section, we first introduce the two major themes of email surveys that measure overall user satisfaction as well as collect feedback on individual product areas on LinkedIn. Then we discuss the survey requirements that ultimately lead to designing and implementing a large-scale online survey framework.

*4.1.1 Two Major Themes: rNPS and MoT.* We run two major themes of NPS surveys. The first theme, called member relationship NPS (rNPS), focuses on long-term customer relationship. In rNPS surveys, we ask LinkedIn members the standard NPS question and collect feedback on several verbatim questions. This survey is an on-going program where we closely monitor the trend for many user segments.

The other theme, called Moment of Truth (MoT) survey, focuses on user experience about specific products. These surveys target members who have just performed certain actions (e.g. searching for people, commenting on feed) or have just received certain experience (e.g. receiving an invitation from another member) on LinkedIn. The NPS question is on the relevant product and the goal is to get timely feedback and to calibrate product development. We care more about targeting the correct audience and having enough survey samples, as some of the experience might be rare. In addition, there are multiple MoT surveys running simultaneously and the sampling and reporting process needs to be continuous over time.

Table 1 summarizes the characteristics of the two major themes.

**Table 1. Comparison of the two major themes of email surveys at LinkedIn**

|  | **rNPS** | **MoT** |
|---|---|---|
| *Population* | All active members in last 180 days | All members performed particular actions or received certain experience during the month |
| *Survey Focus* | General experience on LinkedIn | Particular product experience (e.g. search functionality) |
| *Sampling Frequency* | Daily | Daily for every MoT program |
| *Sample Size* | Large | Small |

*4.1.2 Survey Constraints.* To ensure that survey studies draw meaningful conclusions while minimizing impact on members, the following survey constraints should be satisfied.

**Collect representative samples**. To remove sampling bias, the sampled members should be representative of the population of interest. Other than ignoring sampling bias or trying to adjust for it later, it is best to eliminate it with proper sampling design.

**Sample continuously**. Survey results can be affected by many controllable or uncontrollable external factors, such as big product changes or major world events. Collecting all samples at once increases the vulnerability to such influence and is less likely to get representative samples. In addition, MoT requires that we survey users shortly after their actions. It can only be achieved if we sample continuously.

**Enforce adequate survey cool-off**. Even though we try to be less interruptive the way we ask members to take surveys, the nature of surveying distracts members from their regular activity flow. As a result, it is essential that we enforce adequate cool-off periods between consecutive surveys members receive. For email surveys, we set the cool-off constraints as follows: a member is only eligible to receive a survey if he or she has not been surveyed in the past 30 days, or past 90 days for the same survey program.

**Ensure flexibility and consistency**. The process has to be flexible to accommodate multiple survey programs. In addition, we need to ensure results are consistent over time for existing programs as we continue to add new programs.

As the number of survey programs increases, it becomes harder to satisfy the above constraints and requirements. In Section 4.2 and 4.3, we discuss the rigorous and easy-to-manage survey sampling framework that accommodates the above constraints with the support for multiple survey programs.

### 4.2 rNPS Survey

Relationship NPS survey is one of the most important survey programs at LinkedIn. With rNPS, we monitor our overall NPS trend over time and calibrate product development accordingly.

Simple random sampling is a simple and widely adopted sampling approach. We first introduce the potential biases caused by using simple random sampling, then present our proposed sampling framework.

*4.2.1 Simple Random Sampling for rNPS Surveys.* In our context, if we were to adopt simple random sampling, we would follow the steps below:
1. Randomly select a group of users from all LinkedIn members. The selected members form the rNPS member pool and are ineligible for other email surveys. On the other hand, the unselected members will not receive rNPS surveys but they may receive other surveys such as MoT.

2. In the rNPS pool, determine rNPS eligible members.
3. A fixed percentage of members are then randomly drawn from the rNPS eligible group.

This guarantees no interference between rNPS and other email surveys. If we do not enforce the 90-day same survey cool-off constraints, every member has an equal chance of being selected for rNPS. However, with the cool-off constraints, this is no longer the case. The eligibility for rNPS is based on member's active level. Heavy members appear in the survey population more often than light members and they are more likely to be sampled into rNPS surveys and trigger cool-off. As more and more heavy members become ineligible for rNPS due to cool-off, the samples bias towards light members.

*4.2.2 Pre-Allocated Bucket Sampling.* Simple random sampling creates sampling bias and this bias needs to be addressed during analysis. However, member's visitation pattern cannot be perfectly predicted, and thus, adjustment for such bias is nearly impossible. Therefore, we propose to use pre-allocated bucket sampling (PABS) by leveraging LinkedIn experimentation platform to solve the potential bias caused by the simple random sampling design.

**Experimentation Traffic Allocation System**

The experimentation traffic allocation system at LinkedIn [26] takes hashID as an input to MD5 based algorithm to randomize members onto the interval [0,100). Then, using the randomized value, assign members to treatment or control. Given hashID and memberID, the randomization result is deterministic. Let $f(h, i)$ be such randomization result, where $h$ is the hashId and $i$ is a member's memberID. Assuming a fixed hashID is used, we can omit $h$ and use $f(i)$ instead.

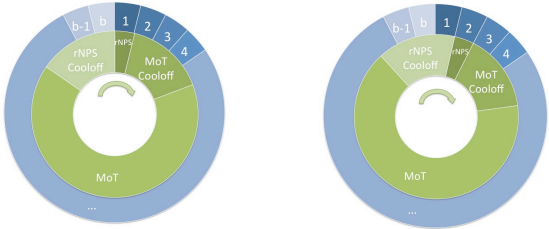

**Figure 1. Demonstration of PABS. Left): Sampling scheme on Day 1; Right): Sampling scheme on Day 2**

**PABS**

On a high level, PABS utilizes randomization from $f(i)$ to construct groups of members eligible for each survey theme. The groups rotate to satisfy cool-off contraints and enable NPS monitoring. The detailed steps are as follows:

1. With $f(i)$, divide all LinkedIn members into $b$ ($b \gg 90$) buckets of $\left[0, \frac{100}{b}\right), \left[\frac{100}{b}, \frac{200}{b}\right), \ldots, \left[\frac{100(b-1)}{b}, 100\right)$ (refer to the blue circle in Figure 1). The exact value of $b$ is set such that sufficient daily samples can be draw from one bucket. The group that a member $i$ falls into, $B(i)$, is determined by

$$B(i) = \left\{ j \mid \frac{100(j-1)}{b} \leq f(i) < \frac{100j}{b}, j = 1, \ldots, b \right\}$$

2. Assign the first bucket to rNPS surveys. Set aside its cool-off group. Similarly, the remaining buckets are for MoT surveys, followed by MoT cool-off group. The cool-off group spans no less than 30 buckets according to the cool-off constraints. Refer to the green circle in Figure 1 Left for the four groups.
3. On Day 1, send rNPS surveys to all eligible members in bucket 1 (Figure 1 Left).
4. On Day 2, rotate the four groups clockwise by one bucket. Send rNPS surveys to eligible members bucket 2 (Figure 1 Right).
5. Similarly, every day the four groups rotate clockwise by one bucket, and a new bucket of users are selected for rNPS surveys.

Compared with simple random sampling, PABS has several good properties:

1. Every day, a new bucket of users are available for rNPS. Daily samples are obtained from an identically distribution and independent of samples drawn on other days. This is one major advantage over simple random sampling.
2. With PABS, samples are drawn continuously. Survey results are less likely to be skewed by particular incidents.
3. Cool-off constraints are automatically ensured. If $b$ is large, by the time a bucket is sampled for rNPS the second time, the cool-off period has passed.

Although in our case we have two survey themes, PABS can be applied to other survey studies even when there is only one or multiple survey programs.

## 4.3 MoT Surveys

Now we switch our attention to MoT surveys. MoT surveys require one to perform certain actions or receive certain experience to trigger the corresponding MoT survey. The triggering rate can be low and we cannot predict when members trigger the required experience. If we adopt the bucket sampling as in rNPS, one bucket of members daily may not yield enough responses. We therefore decide all MoT surveys sharing the sampling pool (see the green circle in Figure 1). Although the MoT group rotates every day, with a large $b$, the day-to-day difference in MoT sampling pool is negligible.

We first illustrate the issues caused by simple random sampling, and then present our proposed solution and the trade-offs made. To simplify the comparison, we assume there is only one type of MoT survey.

*4.3.1 Simple Random Sampling for MoT Surveys.* To adopt simple random sampling, every day, we would sample a fixed percentage from members who triggered the day before, subject to cool-off constraints.

- **Pros**: Daily survey sent volume is fairly consistent with slight fluctuation due to day of week effect (see Figure. 2 Left). As a result, it better captures member's monthly satisfaction towards the product and is more robust against rare incidents that may impact survey results on a particular day.

- **Cons**: Members in the sampling frame have unequal sampling probabilities (see Figure. 3 Left). The more frequently a member triggers, the higher chance he/she is selected. Assume a member triggers $n$ days during the month and every day we sample $r\%$ from the triggered members. The probability of not being chosen at all is $(1 - r\%)^n$ and thus, the probability of being selected is:

$$1 - (1 - r\%)^n \approx nr, if\ r\ is\ small$$

As we can see, frequently triggered members have higher chance to be sampled.

If simple random sampling were to be used, the sampling bias had to be corrected before drawing meaningful conclusions. However, to correct this bias, we need to study individual member's triggering pattern, which is a costly task.

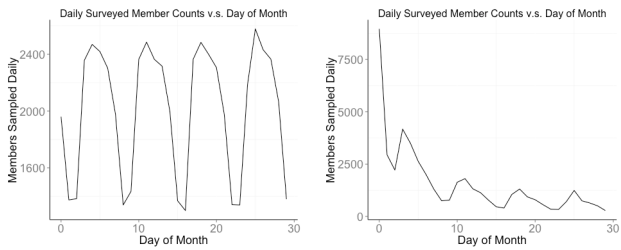

**Figure 2. Daily survey volume with respect to day of month for Left): simple random sampling and Right): FTT sampling.**

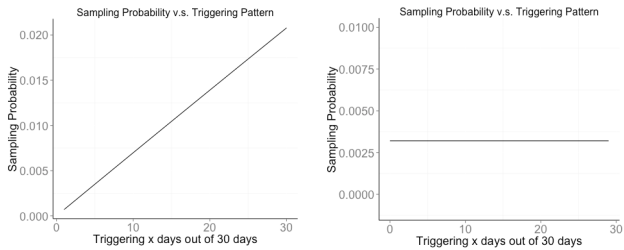

**Figure 3. Sampling probability with respect to member's triggering pattern for Left): simple random sampling and Right): FTT sampling.**

*4.3.2 Sampling from First-Time Triggered Members.* Instead of sampling from all triggered members, we can sample a fixed percentage from eligible first-time triggered (FTT) members in a month; in other words, a member can only be selected the first time he/she triggers in a month. Here are the pros and cons of FTT:

- **Pros**: Members in the sampling frame have an equal chance of being selected (see Figure. 3 Right). Without sampling bias, analyzing the result becomes considerably simpler and more straightforward.
- **Cons**: Majority of survey quota will be fulfilled in the first few days or week (see Figure. 2 Right). More active members trigger more often than less active members and they are more likely to be sampled in the beginning. Less active members tend to trigger later in the month. Because we fulfill most of quota in the beginning, surveys are not sent out evenly in the month. This sampling approach is also vulnerable to abnormalities on the site or breaking events in the first few days of a month.

*4.3.3 Comparison of Two Sampling Methods.* Unbiased sampling eliminates the need of sampling bias adjustment and ensures that our survey study generates trustworthy conclusions. On the other hand, if a breaking event does happen in the beginning and we believe it will impact survey results, we can exclude these days. FTT sampling is preferred because the sampling process is unbiased.

*4.3.4 Members Eligible for Multiple Surveys.* Members can trigger multiple experiences on a single day. For example, a member can comment on the feed and apply for a job on the same day and thus trigger two MoT surveys. Sending multiple surveys to such members violates the sampling constraint of no more than one MoT survey per 30 days. For such members, we need to decide which MoT survey he/she should be assigned to so that the benefit to the overall survey study is the greatest.

Below are the proposed sampling steps:

**Step 1**. Determine the sampling rate and the desired number of survey responses for each MoT survey, and draw random samples from first-time triggered members.

**Step 2**. For members selected by one survey only, we send members the corresponding survey.

**Step 3**. For overlapped members, we send only one eligible survey to the member. The probability of assigning to one survey is inversely proportional to the desired sample size. The rationale is that a rarely triggered MoT survey benefits more from an additional survey response.

We illustrate Step 3 with an example. Assume there are two MoT surveys, A and B. Assume $n_a$ and $n_b$ members are selected for A and B respectively, and among them, $n_{ab}$ members are selected for both. Member $i$ is among the $n_{ab}$ overlapped members. The assignment probabilities of the Member $i$ are determined by:

$$\Pr(assign\ i\ to\ A) = \frac{n_b}{n_a + n_b}$$

$$\Pr(assign\ i\ to\ B) = \frac{n_a}{n_a + n_b}$$

As we can see that if $n_a$ is bigger than $n_b$, we are more likely to assign members to Survey B. This is because there are not as many members available for Survey B. An additional sample is more valuable to Survey B than to A.

During the analysis phase, member's NPS in each MoT survey will need to be adjusted by weights. Take Survey A as an example: for members sampled by both A and B and assigned to A, their weights are $(n_a + n_b)/(n_a n_b)$; for members triggered A and assigned to A, their weights are $1/n_a$ instead. The estimated Survey A NPS and its error margins are computed according to the weights.

The weighted sampling and adjustment can be extended to accommodate the situation when a member is selected for more than two MoT surveys. The sampling and adjustment method adopted here can be applied to any multiple survey studies where survey populations overlap but user can receive no more than one survey.

## 5 IN-PRODUCT SURVEYS

Email surveys are good at capturing the trend of user relationship but in most cases, they are not closely tied to any product iterations. Unlike email surveys, in-product surveys are conducted while users are actively engaging with the product itself. The responses collected are better reflection of users' opinions towards the particular product in use, and hence is more sensitive to product changes. As a result, one big advantage of in-product surveys is that we can closely couple them with new product launches, and use them to compare the change in users' perceptions. At the same time, in-product surveys pose additional challenges compared to email surveys:

*Sampling challenges.* Unlike email surveys where we have full control of whom to sample, in-product surveys are only available to those who use the product during the sampling period. The cohort of survey eligible users for each iteration can vary dramatically due to product changes. Even for the same iteration, the cohort can change over time.

*Nonresponse bias challenges.* Response rates can change from iteration to iteration. Especially when a particular group of users have strong opinions towards the product changes, they are more likely to respond.

In this section, we use a survey study at LinkedIn comparing two mobile apps to further elaborate the above challenges. We then discuss how sampling and bias adjustment is done to overcome these challenges.

### 5.1 NPS Surveys on Two Mobile Apps

In November 2015, LinkedIn launched a new flagship mobile app (codename "Voyager") as the successor for the old app (codename "Titan"). Voyager is a complete architectural and user interface overhaul that aimed to simplify the app with stickier experience. An important question we want to answer is how LinkedIn members perceive Voyager vs. Titan in terms of NPS. The comparison can be formulated as a standard A/B testing problem. We randomly assign member $i$ to use Titan or Voyager app and collect their NPS $X_i$. Individual NPS from promoters, passives and detractors are coded as 100, 0 and -100.

$$X_i = \begin{cases} X_i(t), \text{if } i \text{ assigned to Titan} \\ X_i(v), \text{if } i \text{ assigned to Voyager} \end{cases}.$$

The estimator for Average Treatment Effect (ATE) of the new app can then be estimated by

$$\hat{\Delta} = \frac{1}{N_V} \sum_{\substack{i \text{ assigned} \\ \text{to Voyager}}} X_i - \frac{1}{N_T} \sum_{\substack{i \text{ assigned} \\ \text{to Titan}}} X_i$$

where $N_V(N_T)$ denotes the number of members in Voyager (Titan), assuming stable unit treatment value assumption (SUTVA) and Ignorable Treatment Assignment Assumption [24].

However, the A/B testing formulation has several constraints. As we mentioned earlier, one big challenge with in-product surveys is that we can only sample those who use the product during the survey period. In the Voyager vs. Titan comparison, because the release of the new app is controlled by app stores [25], we cannot truly randomize users into different versions of our flagship app. What's more, even if the sampling process is randomized, voyager-eligible users may not upgrade to the voyager app. As a matter of fact, 12 months after Voyager launch, we have about 5% of users who are still on Titan. These non-adopters tend to be quite different from adopters in terms of both product engagement and NPS responses.

Because of these constraints, we decided to use the pre-post approach where we collected NPS samples on Titan prior to Voyager launch and then on Voyager post launch. While we were able to sample among all active app users for the Titan survey, we were only able to reach the adopters for the Voyager survey. There is clearly a sample bias we need to address.

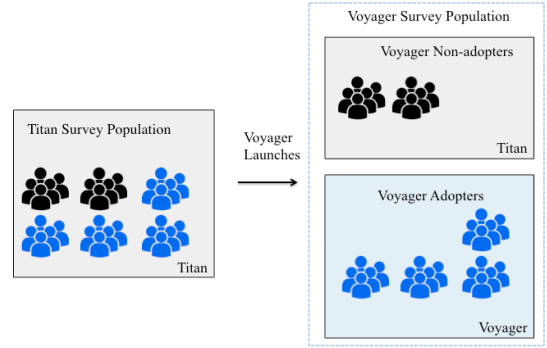

**Figure 4. Comparison of Titan and Voyager NPS Survey Population**

To overcome this challenge, we define the population in the Titan and Voyager NPS study to be *all* mobile app users in the past 6 months before the beginning of the survey (see Figure 4), denoted by $P(t)$ and $P(v)$ respectively. $P(v)$ includes both adopters and non-adopters. The comparable populations allow for constructing unbiased NPS comparison. Users in various countries and markets can react differently to the new app. Let $Y_c(t)$ and $Y_c(v)$ be the population NPS if all app users in country $c$ use Titan or Voyager respectively. Suppose an individual user $i$'s NPS towards Titan is denoted by $X_i^c(t)$. Then $Y_c(t)$ is defined as

$$Y_c(t) = \frac{1}{N_c} \sum_{\substack{user\ i\ in \\ P_c(t)}} X_i^c(t)$$

for country $c$ in $1, \ldots, C$, where $N_c$ is the population size. In particular, we are interested in the change in NPS in $C$ countries of interest.

$$\Delta_c = Y_c(v) - Y_c(t), \text{ for country } c \text{ in } 1, \ldots, C.$$

We estimate $Y_c(t)$ and $Y_c(v)$ by $\widehat{Y_c}(t)$ and $\widehat{Y_c}(v)$ through survey sampling. The estimate of $\Delta_c$, $\widehat{\Delta_c}$, can be obtained via:

$$\widehat{\Delta_c} = \widehat{Y_c(v)} - \widehat{Y_c(t)}, \text{ for country } c \text{ in } 1, \ldots, C.$$

To capture a reasonable number of both light and heavy members, we sampled for a consecutive 30 days on Titan and Voyager separately and collected 2,000 to 4,000 responses in each country of interest. During fielding, for each country $c$, a random group of $p_c\%$ users are eligible to take the survey, where $p_c$ was determined based on the estimated response rate in country $c$. Users can see the survey link at most once regardless whether they take actions or not. This is to ensure good user experience and contain impact on other product areas.

## 5.2 Biases in In-Product Survey Studies

In section, we discuss the two main sources of bias in the context of our case study and in-product survey in general. Both are important and need to be addressed at sampling and analysis stages.

*5.2.1 Sampling Bias.* As we discussed in Section 5.1, we cannot sample non-adopters to take the voyager survey. In addition, the sampling process captures monthly active users while the population we want to study is active users in 6 months. Since users who are more active are more likely to be sampled, this discrepancy contributes one source of sampling bias. Moreover, due to seasonality, users we sampled during pre-Voyager month can be different from users we sampled during the post-Voyager month. Also the system and infrastructure that support survey studies can fail and cause bias.

*5.2.2 Nonresponse Bias.* Now we shift our focus to nonresponse bias. Nonresponse bias occurs when answers of respondents differ from the potential answers of those who did not answer. In a deterministic view, we can split the population into two groups, respondents and nonrespondents. Let $r$ be the response rate. Suppose the average responses are $\mu_r$ and $\mu_n$ for respondents and nonrespondents respectively. The nonresponse bias is given by:

$$\text{Nonresponse Bias} = \text{population mean} - \text{respondent mean}$$
$$= [r\mu_r + (1-r)\mu_n] - \mu_r = (1-r)(\mu_n - \mu_r)$$

Therefore, survey results are more vulnerable to nonresponse bias when the difference between respondents and nonrespondents is larger or the response rate is lower.

**Table 2. Average NPS and response rate in US, by job seeking status.**

| Job Seeking Status | Response Rate | NPS Difference (Active – Inactive) |
|---|---|---|
| **Active** | 0.96% | 22.1 |
| **Inactive** | 0.16% | |

In our study, the overall response rate is around 0.6%. Moreover, the response rates and the actual responses vary among different groups. As a matter of fact, members who look for jobs more actively on LinkedIn are more likely to respond and they respond very differently from their counterparts (see Table 2).

## 5.3 Bias Adjustment

Knowing that both sources of bias exist, we now discuss how we adjust to make proper comparisons between Titan and Voyager. In this section, we explore two popular bias adjustment methods and compare them with before-adjustment. We also discuss why we pick one method over the other for ongoing in-product survey studies.

*5.3.1 Weighting Adjustment.* To adjust for sampling and nonresponse bias, one approach is weighting adjustment method [16]. It assumes that members who share similar characteristics would respond similarly, regardless whether they actually responded to the survey or upgrade. In this approach, we first stratify all users into subgroups. The estimator for the population NPS is then a weighted average of responses from each subgroup:

$$\widehat{Y}_c(A) = \frac{1}{\sum_{k,c} w_{k,c}} \{\sum w_{k,c} \widehat{Y_{k,c}}(A)\},$$

$$k = 1, 2, \dots, K \text{ subgroups}, A = v \text{ or } t$$

where $w_{k,c}$ is the population weight of subgroup $k$.

The challenge with this approach is that $w_{k,c}$ is usually unknown, as the population information is not accessible. Luckily, this is not a problem for us as we keep a set of static attributes for all our members, which can be used to compute the population weights $w_{k,c}$. Notice that the weighting adjustment approach not only adjusts for nonresponse bias, but also corrects the sampling bias because the weights $w_k$'s are obtained from the population.

*5.3.2 MRP.* Another approach, multilevel regression with poststratification (MRP), utilizes propensity models [17, 18, 19]. First, it fits a multinomial regression model using the variables. Then with the fitted response as weights in the population, it computes the adjusted response using post stratification. The model looks as follows:

*Promoter Propensity:*
$$p(\mathbf{z}_i^c) = Pr(X_i^c(A) = 100) = logit^{-1}(\beta_0 + \boldsymbol{\beta}^c \mathbf{z}_i^c),$$

*Detractor Propensity:*
$$q(\mathbf{z}_i^c) = Pr(X_i^c(A) = -100) = logit^{-1}(\alpha_0 + \boldsymbol{\alpha}^c \mathbf{z}_i^c).$$
$$A = v \text{ or } t$$

where $\mathbf{z}_i^c = (z_1, z_2, \dots, z_s)$ captures member $i$'s values among $s$ variables. Each component of $\boldsymbol{\beta}^c$ (or $\boldsymbol{\alpha}^c$), $\beta_j^c, j = 1, \dots, s$, is assumed to be normally distributed with

$$\beta_j^c \sim N(\mu_j^c, \sigma_j^c)$$

where $\sigma_j^c$ is estimated from uninformative prior density function. The estimated NPS is obtained by:
$$\widehat{Y}_c(A) = \frac{1}{N_c} \sum (p(\mathbf{z}_i^c) - q(\mathbf{z}_i^c))$$

*5.3.3 Variables for Adjustment.* Given the population of interest are members on LinkedIn, we have some knowledge of their attributes and behaviors. The list of the initial variable pool is selected to cover various aspects of member's static attribute as well as their engagement with LinkedIn. Nevertheless, the initial pool has over 20 variables. It is noted that for both methods, the more variables introduced, the higher the variance of the estimates due to overfitting.

In the weighting method, since we cross-tabulate using variable combinations, if an excessive number of variables are used, some combination may not have enough samples to represent. We keep two variables maximum to maintain a good number of responses across combinations. The steps below, identical for every country and Titan/Voyager, summarize how we picked these two variables:

1. For every possible variable combination, calculate the NPS estimate $\hat{Y}_v$ given by weighting adjustment.
2. Select the variable combination that is the most effective at adjusting the score, i.e. the combination that gives the most before and after difference.

The variables picked for Voyager and Titan separately turns out to be identical across all the countries. This gives us confidence that our procedure does not inflate the before-and-after adjustment differences. In addition, for almost all countries, "job seeking status" and "active level on LinkedIn" are the two variables picked. For the few countries where other attributes are chosen, the extra gain of using these different variables is small. For easier interpretation and maintenance, we decided to use these two variables for all countries across Titan and Voyager.

In MPR, the weight for each subgroup is deterministic once the population is defined. We start with all the 20 variables in the model and use stepwise logistic regression for variable selection for promoters and detractors. In the final regression model, three variables are selected. "is premium member" is the third variable selected in addition to the same two selected by weighting adjustment.

*5.3.4 Comparing Weighting Adjustment with MRP.* Both methods use population weights for adjustment. MRP allows more variables than the weighting adjustment method. It is not unusual to have very few or no data points for some subgroups in the sample, especially when a large number of variables are used to construct the subgroups. Weighting method would require one to impute the data points for the missing subgroups. However, with a fitted model from MRP, a subgroup's propensity can be estimated, even if no data point is observed for the subgroup. In addition, MRP can use continuous variables directly in the regression model, while in weighting adjustment, a continuous variable has to be bucketized first.

**Table 3. Comparison of NPS adjustments produced by weighting vs. MRP methods.**

| Country | MRP adjustment | Weighting adjustment | Diff | Error Margin |
|---|---|---|---|---|
| Germany | -11.5 | -10.9 | -0.7 | 6.8 |
| India | -2.3 | -0.6 | -1.7 | 4.1 |
| US | -6.4 | -6.5 | -0.1 | 3.2 |

From Table 3, we see that weighting adjustment and MRP are effective in correcting nonresponse bias (Column 1 and 2). The adjusted NPS from both methods are very close. The difference (Column 3), compared with error margin (Column 4), is negligible. Especially given that the majority of the picked variables for the two methods overlap, we pick the weighting method to correct bias because it is more interpretable.

With the estimated NPS for both Titan and Voyager, we can conclude which app has higher NPS. This comparison uses the two-sample t-test, the details of which can be found in standard statistics textbooks and hence are omitted here.

Although in the case study, we compare surveys results from two mobile apps, the same methodology can be extended to comparisons of results from any two surveys. When comparing results from two surveys, as long as the populations are comparable and bias is corrected, inferences on the populations from individual surveys are comparable as well.

## 5.4 Continuous Sampling in In-product Surveys

As products are constantly iterating and evolving, each new version comes with bug fixes and new features. Collecting meaningful in-product samples and monitoring NPS over time help guide us in product development. In this section, we extend PABS to in-product surveys and connect survey studies with A/B testing.

*5.4.1 PABS in In-product Surveys.* While we could show the survey every time a user uses the product, this experience is intrusive. Moreover, in-product surveys distract users from engaging with the product. Similar to email surveys, enforcing a survey cool-off is essential.

We adopt PABS for in-product surveys with slight modification. The entire LinkedIn member base is divided into $b$ buckets. The bucket where a user falls into is determined by $f(h, i)$. Here the hashID $h$ is different so that the app survey is orthogonal to rNPS and MoT surveys. In Figure 5, we demonstrate the sampling approach with 5 buckets.

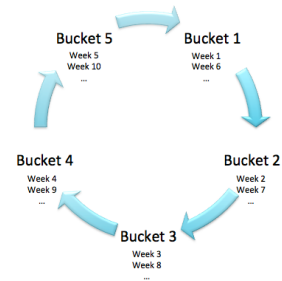

**Figure 5. PABS for mobile app continuous sampling**

- Every week, one bucket of users are eligible to see the survey in product.
- The next week, we show survey to users in the next bucket. The survey eligible bucket rotates weekly.

Every week, we weight-adjust the collected responses and report NPS for that week. The sampling scheme here shares the general benefits of PABS: comparable sampling frame and continuous reporting over time. In addition, it constructs an A/B test that compares survey experience with regular experience.

*5.4.2 A/B Testing on Surveys.* When conducting an in-app survey, besides the usual operational cost from hosting to data collection, there is another type of cost that is often overlooked - the opportunity cost that other content could have been served to the user instead of survey call to actions. Users could have clicked an article or ad instead of the survey. Further more, the call to action pushes other content below and may cause users to scroll less likely and consume less content.

By constructing the A/B test, we can measure how surveys affect user's activities on LinkedIn and the downstream monetary impact. This, in return, provokes us to think about the impact on Scalable Online Survey Framework: from Sampling to Analysisthe whole LinkedIn ecosystem when running surveys. In the A/B tests constructed by our survey studies, we have observed that survey experience indeed causes consistent short-term business metric drops. However, with proper cool-off constraints, the metric impact goes away in the medium and long term.

## 6 CONCLUSIONS AND FUTURE WORK

In this paper, we discussed many interesting and widely applicable survey-related questions at LinkedIn. We formulated and established a scalable survey framework that can be easily extended to solve many similar issues. We studied survey sampling and analysis in the context of email and in-product surveys separately. We also briefly discussed using A/B test to quantify the product impact from survey studies.

One interesting topic that we did not explore much is how to validate the bias adjustment in Titan vs. Voyager study. We try to remove bias as much as we can, but in reality, we do not know how much bias is left. The bias-free estimate is often non-trackable unless we reach the nonrespondents via other methods. The comparison study of Titan vs. Voyager is similar to running an online A/B test, except that the time factor was not controlled. If a user's app version can be randomized, an easy comparison can be made through A/B testing.

We use email as the primary channel for rNPS and MoT surveys because sending emails is cheap. However, email is not the best channel to collect immediate user feedback especially for MoT surveys. We are switching to in-product trigger-based MoT survey. When a user interacts with LinkedIn products and triggers a particular survey, an in-product questionnaire will pop up and ask for user's immediate feedback. We believe the switch to in-product survey will greatly increase the response rate and collect first-hand user feedback.